\documentclass[aps,prb,english,footinbib,floatfix,reprint,twocolumn,showpacs,superscriptaddress,longbibliography,10pt,tightenlines]{revtex4-1}
\usepackage[utf8]{inputenc}
\usepackage{amssymb,amsmath}
\usepackage{graphicx}
\usepackage{natbib}
\usepackage{multirow}
\usepackage{physics}
\usepackage{subcaption}
\usepackage{ragged2e}
\DeclareCaptionJustification{justified}{\justifying}
\usepackage[final]{hyperref} 
\hypersetup{
	colorlinks=true,       
	linkcolor=red,        
	citecolor=red,        
	filecolor=magenta,     
	urlcolor=blue         
}

\usepackage[dvipsnames]{xcolor}

\begin{document}

\title{Non-equilibrium time evolution in the sine-Gordon model revisited}

\author{D. Sz{\'a}sz-Schagrin}
\affiliation{Department of Theoretical Physics, Institute of Physics,\\ Budapest University of Technology and Economics, H-1111 Budapest, M{\H u}egyetem rkp. 3}
\affiliation{BME-MTA Momentum Statistical Field Theory Research Group, Institute of Physics,\\ Budapest University of Technology and Economics, H-1111 Budapest, M{\H u}egyetem rkp. 3}
\author{I. Lovas}
\affiliation{Kavli Institute for Theoretical Physics, University of California, Santa Barbara, CA 93106, USA}
\author{G. Tak{\'a}cs}
\affiliation{Department of Theoretical Physics, Institute of Physics,\\ Budapest University of Technology and Economics, H-1111 Budapest, M{\H u}egyetem rkp. 3}
\affiliation{BME-MTA Momentum Statistical Field Theory Research Group, Institute of Physics,\\ Budapest University of Technology and Economics, H-1111 Budapest, M{\H u}egyetem rkp. 3}
\affiliation{HUN-REN–BME Quantum Dynamics and Correlations Research Group, Institute of Physics,\\ Budapest University of Technology and Economics, H-1111 Budapest, M{\H u}egyetem rkp. 3}
\date{31st August 2023}

\begin{abstract}
We study the non-equilibrium dynamics of the quantum sine-Gordon model describing a pair of Josephson-coupled one-dimensional bosonic quasi-condensates. Motivated by experimentally accessible quench procedures where the zero mode of the quasi-condensates is weakly coupled to finite momentum modes, we develop a novel Hamiltonian truncation scheme relying on a mini-superspace treatment of the zero mode (MSTHA). We apply this method to simulate the time evolution after both weak and strong quantum quenches, injecting a low or high energy density into the system, and demonstrate that MSTHA accurately captures the dynamics from the hard core boson limit to the experimentally relevant weakly interacting regime for sufficiently mild quenches. In the case of high energy densities, MSTHA breaks down for weak interaction but still extends the range of validity of previous Hamiltonian truncation schemes. We also compare these results to the semiclassical truncated Wigner approximation (TWA) and establish that the dynamics can be well approximated by the semiclassical description in the weakly interacting regime realised in the experiments. In addition, we clarify the importance of the phononic modes depending on the sine-Gordon interaction strength.
\end{abstract}

\maketitle

\section{Introduction}

The sine-Gordon model is a paradigmatic example of integrable quantum field theories and also an effective description of the low-energy physics of numerous physical systems, such as, e.g., spin chains\cite{Affleck1999, Umegaki2009, Zvyagin2004}, circuit quantum eletrodynamics\cite{Roy2019, Roy2021}, and bosonic and fermionic Hubbard models\cite{Essler2005book, Giamarchi2004book, Nagerl2010, Controzzi2000}. Due to its integrability, many equilibrium properties of the model are exactly known, ranging from exact results on scattering amplitudes and form factors to expectation values of local observables\cite{Zamolodchikov1978,Lukyanov1997,Smirnov1992,Essler2005book,Hegedus2018,Buccheri2014, Horvath2018}.

Recently, it also attracted considerable interest in the context of non-equilibrium dynamics due to an experimental realisation with two Josephson-coupled one-dimensional bosonic quasi-condensates\cite{Hofferberth_2007,Langen_2013,Gring_2012,Tajik2022, Bouchoule2005}. In the experiment, ultra-cold atoms are trapped in an elongated double-well potential, limiting the physics to one spatial dimension. The effective description of the system can be obtained using bosonisation\cite{Gritsev2007}, predicting that the anti-symmetric modes of the double-well potential realise the sine-Gordon model, weakly coupled to a Luttinger liquid accounting for symmetric modes. These considerations suggested new possibilities for experimentally observing the out-of-equilibrium dynamics of the sine-Gordon model, an idea gaining further experimental support by demonstrating that correlations in thermal equilibrium can be described in terms of the (classical) thermal sine-Gordon.\cite{Schweigler2017,Beck2018}. However, non-equilibrium phenomena observed in the experiments point to dynamics beyond the sine-Gordon model\cite{Pigneur2018,Horvath2019,2021PhRvR...3b3197M,2021ScPP...10...90V}. These results show the relevance of coupling terms to additional degrees of freedom, such as the symmetric modes or the transverse modes in the quasi one-dimensional geometry. Identifying the simplest theoretical model accounting for the experimental observations remains an outstanding open question, the resolution of which requires efficient numerical methods that yield reliable predictions for the experimental protocols.

In parallel with and motivated by these experimental developments, several theoretical approaches have been developed along different lines of approach. A paradigmatic and experimentally relevant framework to out-of-equilibrium dynamics in quantum many-body systems is provided by the framework of quantum quench\cite{Calabrese2006,Calabrese2007}. In this scenario, the system is initially in equilibrium, prepared in the ground state of some pre-quench Hamiltonian. It is then driven out of equilibrium by a sudden change of some parameters, leading to a subsequent evolution governed by a different (post-quench) Hamiltonian. Several avenues can be explored to describe the time evolution of the sine-Gordon model after a quantum quench. These include semiclassical approximations, such as the mean-field approximation\cite{Nieuwkerk2019,Nieuwkerk2020} or the truncated Wigner approximation (TWA)\cite{Polkovnikov2003, POLKOVNIKOV2010, 2013PhRvL.110i0404D, Horvath2019}. However, semiclassical approaches are, in general, uncontrolled approximations that need to be validated against some complementary description of the quantum dynamics to test their validity. An alternative is a form factor expansion relying on the exactly known spectrum and local operator matrix elements (form factors) of the model\cite{Bertini2014,2017JSMTE..10.3106C}; however, this runs into serious difficulties in the experimentally relevant attractive regime\cite{Horvath2018}. Another way to describe non-equilibrium behaviour is provided by generalised hydrodynamics \cite{2016PhRvL.117t7201B,2016PhRvX...6d1065C}, the application of which needs an effective description of thermodynamic states in the model which was resolved only very recently \cite{2023arXiv230316932K,2023arXiv230515474N}.

In this work, we consider an alternative approach to non-equilibrium dynamics provided by the framework of Hamiltonian truncation (THA), a family of numerical approaches to low-dimensional quantum field theories. It was initially developed to describe relevant perturbations of simple conformal field theories \cite{Yurov_1989} and later extended to the sine-Gordon model \cite{Feverati_1998}. Recently it was applied to describe non-equilibrium time evolution, both in perturbed minimal conformal field theories \cite{Rakovszky_2016}, the sinh-Gordon model\cite{Konik2021JHEP} and in the sine-Gordon model 
\cite{Horvath2019,Horvath2022in}; however, previous approaches were limited to parameters away from the experimentally available range. Aiming at overcoming this difficulty, here we introduce a novel truncated Hamiltonian formulation of the sine-Gordon model, which makes use of the so-called minisuperspace approach originally introduced in the context of $\varphi^4$ field theory\cite{Rychkov_2016,Bajnok_2016}. The main idea behind this approach is that in the limit of large Luttinger parameter $K$ relevant for the experiment, the coupling of the zero mode to the non-zero modes is weak. Therefore, solving the zero mode in a numerically exact way and including the non-zero modes afterwards is natural. We compare the results of the new minisuperspace-based truncated Hamiltonian approximation (MSTHA) to TWA for verification and testing the conditions and the range of validity for both approaches.

We find that the MSTHA is well suited for simulating mild quantum quenches inserting a small energy density. For these protocols, MSTHA allows us to obtain reliable, well-converged results even in the experimentally relevant weakly interacting limit, a regime inaccessible by previous implementations of truncated conformal space approach (TCSA). In contrast, for stronger quenches, MSTHA continues to show good convergence properties in the limit of strong interactions but breaks down with decreasing interaction strength, a limitation similar to the one observed in previous TCSA simulations. In contrast to TCSA, semiclassical approaches are expected to become more reliable for stronger quenches or weaker interactions. In accordance with these general expectations, we find that TWA yields a considerable error for weak quenches in the limit of strong interactions, considerably overestimating the damping of quantum oscillations. However, the performance of TWA improves rapidly with decreasing interaction strength, and TWA shows excellent agreement with the essentially exact MSTHA results for moderate interactions. Similarly, larger quenches with a higher energy density render TWA results more reliable, and a direct comparison with well-converged MSTHA reveals considerable errors only close to the limit of hard-core repulsion. These results establish MSTHA and TWA as powerful complementary approaches for simulating sine-Gordon dynamics. Moreover, by considering the mode-resolved occupation numbers for various quench protocols, we take a step towards identifying the most relevant degrees of freedom for the dynamics, an essential ingredient for constructing a simple theoretical model accounting for experimental observations.

The outline of the paper is as follows. In section \ref{sec:sG} we briefly review the sine-Gordon model, and in section \ref{sec:methods}, we describe the MSTHA and its implementation, together with a brief review of the TWA. Section \ref{sec:results} contains our results regarding the time evolution from two different classes of initial states, corresponding to mild and strong quenches with small and high energy density, respectively, together with a comparison to the TWA description. We discuss the results and draw our conclusions in Section \ref{sec:conclusions}. Some technical details are relegated to the Appendix to make the main exposition easier to follow.

\section{Brief summary of the sine-Gordon model}\label{sec:sG}

The classical sine-Gordon model is defined by the following action,
\begin{equation}
    \mathcal{S}_\textrm{sG}^{\textrm{cl}} = \int dt\int dx\left[\frac{1}{2}(\partial_t\varphi)^2 - \frac{1}{2}(\partial_x\varphi)^2 + \lambda\cos\beta\varphi\right],
\end{equation}
describing the continuum limit of a one-dimensional chain of torsion-coupled pendula. 

It has topologically charged soliton/anti-soliton excitations with mass 
\begin{equation}
    M_\text{cl} = \frac{8\sqrt{\lambda}}{\beta}\,,
\end{equation}
and spatially localised oscillating configurations parametrised by a continuous parameter $\varepsilon$ called breathers with mass
\begin{equation}
    m_\varepsilon = \frac{16\varepsilon\sqrt{\lambda}}{\beta}.
\end{equation}
At the quantum level, the classical field $\varphi$ is replaced by the field operator $\hat{\varphi}$ and its dynamics is governed by the Hamiltonian:
\begin{equation}
    \hat{H}_\textrm{sG} = \int dx:\left(\frac{1}{2}(\partial_t\hat{\varphi})^2 + \frac{1}{2}(\partial_x\hat{\varphi})^2 - \lambda\cos\beta\hat{\varphi}\right):\,,
     \label{sg-ham}
\end{equation}
where the semicolon denotes normal ordering relative to the modes of the $\lambda=0$ massless free boson. The spectrum of the breathers becomes discrete:
\begin{equation}
    m_n = 2 M\sin\frac{\pi\xi n}{2}, \quad \xi = \frac{\beta^2}{8\pi-\beta^2}\,,
\end{equation}
where $M$ is the quantum soliton mass. Integrability allows to determine the exact relation between the mass scale given by, say, the first breather mass $m_1$ and $\lambda$ \cite{Zamolodchikov1995}:
\begin{equation}
    \lambda = \left(2\sin\frac{\pi\xi}{2}\right)^{2\Delta-2}\frac{2\Gamma(\Delta)}{\pi\Gamma(1-\Delta)}\left(\frac{\sqrt{\pi}\Gamma\left(\frac{1}{2-2\Delta}\right)m_1}{2\Gamma\left(\frac{\Delta}{2-2\Delta}\right)}\right)^{2-2\Delta}\label{lambda}
\end{equation}
where 
\begin{equation}
    2\Delta = \frac{\beta^2}{4\pi}
\end{equation}
is the anomalous dimension of the cosine operator. All physical quantities can then be parameterised in units of the mass scale $m_1$. We note that another common parametrization relies on the Luttinger parameter
\begin{equation}\label{eq:K}
    K=\dfrac{\pi}{\beta^2}\,,
\end{equation}
with $K=1$ corresponding to hard-core repulsion between bosons, and $K$ increasing with decreasing sine-Gordon interaction strength, such that $K\to\infty$ upon approaching the non-interacting field theory limit.

In a finite spatial volume $L$, observing that the sine-Gordon field is an angular variable of period $\frac{2\pi}{\beta}$, it is natural to consider the following quasi-periodic boundary conditions:
\begin{equation}
    \hat{\varphi}(x+L,t) = \hat{\varphi}(x, t) + \frac{2\pi}{\beta}m \,,
\end{equation}
with $m\in\mathbb{Z}$ giving the so-called winding number a.k.a. the topological charge. We only consider the sector $m=0$ in the following, so the field satisfies ordinary periodic boundary conditions.

The Hamiltonian can be considered as a perturbation of the compactified massless free boson in finite volume with the Hamiltonian
\begin{equation}
    \hat{H}_\text{FB} = \frac{1}{2}\int_0^L dx:\left[(\partial_t\hat{\varphi})^2+(\partial_x\hat{\varphi})^2\right]:
    \label{cftham}
\end{equation}
Expanding the field in Fourier modes
\begin{align}
    \hat{\varphi}(x,t) = &\hat{\varphi}_0 + \frac{1}{L}\hat{\pi}_0 t\nonumber\\ 
    &+ \frac{i}{\sqrt{4\pi}}\sum_{k\neq0}\frac{1}{k}\left[a_k e^{i\frac{2\pi}{L}k(x-t)}+ \bar{a}_k e^{-i\frac{2\pi}{L}k(x+t)}\right]\label{phase-field},
\end{align}
the free part of the Hamiltonian (\ref{sg-ham}) can be written as
\begin{equation}
    \hat{H}_{\text{FB}} = \frac{2\pi}{L}\left(\frac{\hat{\pi}_0^2}{4\pi} + \sum_{k>0}a_{-k}a_{k} + \sum_{k>0}\bar{a}_{-k}\bar{a}_{k} - \frac{1}{12}\right).\label{ham-cft}
\end{equation}
Here $\hat{\pi}_0$ is the zero mode of the momentum canonically conjugate to $\hat{\varphi}$,
\begin{equation}
    \hat{\pi}(x, t) = \partial_t\hat{\varphi}(x, t);\quad\hat{\pi}_0 = \int_0^L dx \hat{\pi}(x,t)\,,
\end{equation}
while the $a_k$ and $\bar{a}_k$ with negative (positive) $k$ are the left and right bosonic creation (annihilation) operators satisfying the commutation relations
\begin{equation}
    [\hat{\varphi}_0, \hat{\pi}_0] = i;\quad [a_k, a_l] = [\bar{a}_k, \bar{a}_l] = k\delta_{k+l, 0}.
\end{equation}
Therefore the sine-Gordon Hamiltonian (\ref{sg-ham}) takes the form
\begin{align}
    \hat{H}_\text{sG} = \frac{2\pi}{L}\left(\frac{\hat{\pi}_0^2}{4\pi} + \sum_{k>0}a_{-k}a_{k} + \sum_{k>0}\bar{a}_{-k}\bar{a}_{k} - \frac{1}{12}\right)\nonumber\\
    -\frac{\lambda}{2}\int_{0}^{L}:\left(e^{i\beta\hat{\varphi}}+e^{-i\beta\hat{\varphi}}\right):\label{sg-ham-tcsa}.
\end{align}

\section{Simulating the time evolution}\label{sec:methods}

\subsection{Truncated Conformal Space Approach}\label{subsec:TCSA}

The main idea of TCSA is to use the eigenstates of the massless free boson in a finite volume $L$ as a computational basis and truncate it by imposing an upper energy cutoff. Since the matrix elements of the exponential operators can be explicitly computed, the Hamiltonian \eqref{sg-ham} can be represented by a finite matrix, reducing the determination of the spectrum and time evolution of expectation values of observables to a numerical linear algebra problem. However, the results obtained through TCSA differ from the exact results by the so-called \textit{truncation errors}. For relevant perturbations, the truncation errors decrease with increasing energy cutoff, and renormalisation group methods can improve the convergence \cite{2006hep.th...12203F,2007PhRvL..98n7205K,2011arXiv1106.2448G,Rychkov_2015,Konik2021JHEP}. While powerful in general, we found that for the quenches studied here, the application of the renormalisation group improvement did not alter the results and the rate of convergence. For this reason, the MSTHA results presented here do not involve such improvement.

The Hilbert space of the massless free boson consists of Fock modules $\mathcal{F}_\nu$
\begin{equation}
    \mathcal{H}_{\text{FB}} = \bigoplus_{\nu \in\mathbb{Z}}\mathcal{F}_\nu \label{hilbert-cft},
\end{equation}
with
\begin{equation}
    \mathcal{F}_\nu = \left\{\ket{\psi} = \prod_{k>0}{a_{-k}^{r_k}}{\bar{a}_{-k}}^{\bar{r}_k}\ket{\nu}\bigg| r_k, \bar{r}_k\in\mathbb{N}^{+}\right\}
    \label{fock-module}
\end{equation}
built upon zero mode plane wave states defined as
\begin{equation}
    \ket{\nu} = e^{i\nu\beta\hat{\varphi}_0}\ket{0}\label{plane-wave-basis}.
\end{equation}
It is useful to further decompose the Fock modules into different momentum sectors parameterised by a quantum number $s\in\mathbb{Z}$ as
\begin{equation}
      \mathcal{F}_\nu = \bigoplus_{s\in\mathbb{Z}}\mathcal{F}_{\nu}^{(s)}
\end{equation}
where
\begin{equation}
    \mathcal{F}_{\nu}^{(s)} = \left\{\ket{\psi} = \prod_{k>0}{a^{r_k}_{-k}}{\bar{a}_{-k}}^{\bar{r}_k}\ket{\nu}\bigg| \sum k r_k - \sum k \bar{r}_k = s\right\}\,,
\end{equation}
with fixed total spatial momentum $2\pi s/L$. In our simulations, we only need the zero-momentum sector, i.e., $s = 0$. 

The Hilbert space is then usually truncated by introducing an upper limit on the unperturbed energy of the massless free boson basis vectors \cite{Feverati_1998}
\begin{equation}
\begin{split}
    \mathcal{H}_{\text{FB}}^{\text{trun.}} = \text{span}
    \bigg\{&
    \prod_{k>0}{a^{r_k}_{-k}}{\bar{a}_{-k}}^{\bar{r}_k}\ket{\nu}
    \\
    &\bigg|
    \frac{(\nu\beta)^2}{4\pi}+\sum\limits_{k>0} k (r_k+\bar{r}_k)<e_\text{cut}
    \bigg\}\label{naive-trun}
    \,.    
\end{split}
\end{equation}
The disadvantage of this truncation procedure is that for small $\beta$, i.e., in the limit of large $K$ corresponding to a weakly interacting quantum field, it includes a large number of Fock modules, which severely limits the method's applicability in the experimental regime\cite{Horvath2019}.

\subsection{The mini-superspace based THA}\label{sec:MSTHA}

To go beyond the TCSA detailed in the previous subsection, we note that the bosonic field $\hat{\varphi}(x,t)$ can be decomposed into homogeneous (zero mode) and inhomogeneous (oscillator modes) parts
\begin{equation}
    \hat{\varphi}(x,t) = \hat{\varphi}_0(t) + \tilde{{\varphi}}(x,t)\,.\label{field-decomp}
\end{equation}
Neglecting the contribution of oscillator modes, the single mode description of the model describes a quantum pendulum:
\begin{equation}
    \hat{H}_\text{QP} = \frac{1}{2L}\hat{\pi}_0^2 -\lambda L \left(\frac{2\pi}{L}\right)^{2\Delta}\cos(\beta\hat{\varphi}_0)\,,
    \label{qp-ham}
\end{equation}
(for the volume dependence see Appendix \ref{appendix-matelements}). The full sine-Gordon model itself corresponds to a quantum pendulum coupled to a set of non-linear, interacting phononic modes:
\begin{align}
    \hat{H}_{\text{sG}} =\frac{1}{2}\int_0^L:\left[(\partial_t\hat{\varphi}_0)^2 + (\partial_t\hat{\tilde{\varphi}})^2+(\partial_x\hat{\tilde{\varphi}})^2\right]: \nonumber\\- \frac{\lambda}{2}\int_0^L dx:\left[e^{i\beta\hat{\varphi}_0}e^{i\beta\tilde{\varphi}} + e^{-i\beta\hat{\varphi}_0}e^{-i\beta\tilde{\varphi}}\right]:\label{sg-ham-decomposed}
\end{align}
In the experimental parameter regime of weak interactions, $\beta$ is small, so the inter-mode coupling is expected to be weak. As a result, it is reasonable to introduce a different approximation to the dynamics, in which the zero-mode dynamics is first solved in a (numerically) exact way, and the coupling to the non-zero modes is taken into account at the next stage, which is known as the mini-superspace approach\cite{Rychkov_2016,Bajnok_2016}. The usefulness of this approach can also be understood by looking at the truncated Hamiltonian approximation as a variational method: optimizing the variational basis allows for more precise computation of spectral quantities and expectation values. 

The first step consists of constructing the single (zero) mode Hamiltonian (\ref{qp-ham}) (the quantum pendulum) on the plane wave basis $\{\ket{\nu}\}$ \eqref{plane-wave-basis} with some appropriate truncation. Diagonalisation of \eqref{qp-ham} yields the energy spectrum and eigenvectors of the pendulum
\begin{equation}
    \hat{H}_\text{QP}\ket{n} = \varepsilon_n\ket{n} \quad n\in\mathbb{N}
\end{equation}
as a function of the truncation of the basis $\{\ket{\nu}\}$. With a high enough truncation, it turns out that the energy levels converge very fast to an essentially exact result.

In the next step, one computes a numerically exact matrix representation of the operators $\hat{\pi}_0^2$ and $e^{\pm i\beta\hat{\varphi}_0}$ on the eigenbasis $\{\ket{n}\}$. The matrix elements of the non-zero mode parts can be computed separately, and their handling can be made more efficient by exploiting the factorisation of the oscillator modes into left- and right-moving sectors. This reduces the memory requirements of the method and enables higher truncation values, similar to the chirally factorised TCSA developed by Horvath et al.\cite{Horvath2022}. As a final step, the sine-Gordon Hamiltonian can be assembled from the finite zero and non-zero mode matrix pieces according to (\ref{sg-ham-decomposed}) by simple matrix operations.

Truncation now depends on two parameters: $n_\textrm{max}$ describing the truncation of the zero mode space and $\ell_\text{cut}$ giving the truncation of the non-zero modes,
\begin{equation}
\begin{split}
    \mathcal{H}_{\text{FB}}^{\text{trun.}} = \text{span}
    \bigg\{&
    \prod_{k>0}{a^{r_k}_{-k}}{\bar{a}_{-k}}^{\bar{r}_k}\ket{n}
    \bigg|
    n\leq n_\text{max}\,\, \text{and} 
    \\
    & \sum\limits_{k>0} k (r_k+\bar{r}_k)\leq \ell_\text{cut}
    \bigg\}\label{mss-trun}
    \,.    
\end{split}
\end{equation}
Time evolution in the TCSA is computed using the Bessel-Chebyshev method\cite{Rakovszky_2016, Horvath2022}. The validity of the results is maintained through monitoring of the norm of the time-evolved state $\ket{\Psi(t)}$
\begin{equation}
    \ket{\Psi(t)} = e^{i\hat{H}_\textrm{sG}t}\ket{\Psi_0}.
\end{equation}
where the initial state $\ket{\Psi_0}$ depends on the quench protocol. For the quantum quenches considered in this paper, it is specified in the next section.

Before applying the method to non-equilibrium time evolution, the zero mode spectrum was cross-checked by comparing it with a solution of the quantum pendulum Schrödinger equation using the shooting method. In addition, we compared the time evolution of the system truncated to its zero mode to a numerical solution of the coordinate space Schrödinger equation for the time evolution. The fully assembled MSTHA was verified by checking the spectrum against the predictions of the exact $S$-matrix sine-Gordon theory and by comparing it to previous TCSA results for the time evolution for quench protocols where they were available. The convergence of the method can also be checked by comparing results for different values of the cutoff; examples are given in Appendix \ref{app:mstha_convergence}.

In our subsequent simulation of time evolution, we consider two observables:
The expectation value of the cosine of the phase field, 
\begin{equation}
    \expval{:\cos\beta\hat{\varphi}:},
\end{equation} 
and the Fourier transform of the phase-phase correlator
\begin{align}
    &\expval{\hat{\varphi}_k\hat{\varphi}_{-k}} =\nonumber\\
    &\frac{1}{4\pi}\expval{\frac{1}{k^2}\left(a_{-k}a_k + \bar{a}_{-k}\bar{a}_k - a_{k}\bar{a}_k - a_{-k}\bar{a}_{-k} + k\right)}\,.
\label{tcsa-corr}
\end{align}
Both are experimentally accessible observables, the first one characterizing the phase coherence between Josephson-coupled one-dimensional bosonic quasi-condensates, which has already been measured for various quench protocols. The latter gives information on the mode-resolved occupation numbers, allowing us to identify the finite momentum modes that contribute substantially to the dynamics. 

\subsection{Truncated Wigner approximation}

The TWA is implemented using the lattice regularisation of the sine-Gordon model\cite{Horvath2019}
\begin{align}
    \hat{H}_{\text{Lat}} = \frac{a}{2}\sum_{j = 1}^{N}\left((\partial_t\hat{\varphi}_j)^2 + \frac{(\hat{\varphi}_j-\hat{\varphi}_{j-1})^2}{a^2}\right)\nonumber\\
    - \dfrac{\lambda a}{\mathcal{N}}\sum_{j = 1}^{N}\cos\beta\hat{\varphi}_j,
\end{align}
with lattice constant $a=L/N$, and the discretised scalar field variables related to the continuum filed via $\hat{\varphi}_j=\hat{\varphi}(x=ja)$. The canonically conjugate momentum variables are given by
\begin{equation}
\hat{\pi}_j=a\partial_t\hat{\varphi}_j,
\end{equation}
and satisfy $[\hat{\varphi}_j,\hat{\pi}_{j^\prime}]=i\delta_{j,j^\prime}$.
Normal ordering of the cosine operator is accounted for by a coefficient $\mathcal{N}$ determined from the Baker-Campbell-Hausdorff formula,
\begin{equation}
    \cos\beta\hat{\varphi}_i = \mathcal{N}:\cos\beta\hat{\varphi}_i:,
\end{equation}
expressed as\cite{Horvath2019}
\begin{equation}
    \mathcal{N} = \exp{\left(-\frac{\pi\Delta}{N}\right)}\prod_{n = 1}^{N/2-1}\exp{\left(-\frac{2\pi\Delta}{N\sin{\frac{\pi n}{N}}}\right)}.
\end{equation}
The Fourier modes of the discretised scalar field are defined as
\begin{equation}
    \hat{\varphi}_{k\neq0} = \frac{1}{N}\sum_{j = 1}^{N}e^{i\frac{2\pi}{N}kj}\hat{\varphi}_j\,.\label{twa-fourier-app}
\end{equation}
The expectation value of their correlator in the ground state is given by
\begin{equation}
    \expval{\hat{\varphi}_k\hat{\varphi}_{-k}}{0} = \frac{1}{4N \sin(\pi k /N)}\,,
\end{equation} 
reducing to the  correlator (\ref{tcsa-corr}) in the continuum limit $N\to \infty$.

In the TWA, the time evolution of operator expectation values is expressed in terms of the Wigner function, defining a  quasi-probability distribution in phase space:
\begin{equation}
\begin{split}
     &W(\underline{\varphi},\underline{\pi}) = \\
     &\frac{1}{(2\pi)^{2N}}\int d\underline{\varphi}'\expval{\underline{\varphi}-\underline{\varphi}'/2 |\hat{\rho}|\underline{\varphi}+\underline{\varphi}'/2}e^{-i  \underline{\varphi}'\cdot\underline{\pi}}\,.
\end{split}
\end{equation}
Here $\hat\rho$ is the density operator corresponding to the state of the system at $t = 0$, and we have introduced the usual vector notation for phase space coordinates
\begin{equation}
    \underline{\varphi} = \{\varphi_j|j = 1, ..., N\}\, ,\quad   
    \underline{\pi} = \{\pi_j|j = 1, ..., N\}\,.    
\end{equation}
Given an initial state $\ket{\Psi_0}$, the corresponding Wigner function can be computed from the density operator $\hat\rho_{\Psi_0}=\ket{\Psi_0}\bra{\Psi_0}$. The TWA approximates the time evolution through an ensemble of classical trajectories, obtained by evolving fluctuating initial conditions $\{\underline{\varphi}, \underline{\pi}\}$, distributed according to the Wigner quasi-probability distribution, with the classical equations of motion. In practice, the calculation is performed through classical Monte Carlo averaging. The Wigner function $W$ is often positive semi-definite~\footnote{The TWA can still be implemented in the presence of negative regions in $W$, but it becomes less efficient due to the so-called sign problem.}, allowing to generate a sufficiently large set of random initial conditions $\{\underline{\varphi}, \underline{\pi}\}$ distributed according to $W$. The time evolution of observables is then computed by averaging over the classical trajectories determined by these initial conditions. A detailed discussion of the TWA implementation and parameter matching with the truncated Hamiltonian approximation has been described previously\cite{Horvath2019}, and we refrain from repeating it here.

\section{Time evolution in sine-Gordon quenches}\label{sec:results}

We now turn to the non-equilibrium time evolution after quantum quenches in the sine-Gordon field theory. Setting the energy unit as $m_1 = 1$, we define the dimensionless volume parameter as $l = m_1L$. Time is measured using the variable $\nu_1t$ where
\begin{equation}
    \nu_1 = \frac{m_1}{2\pi}
\end{equation}
is the frequency associated with the rest mass of the lightest breather. Given the relation to the experimental setup discussed in Appendix \ref{app:condensates}, connection with the experiments is facilitated by characterising the strength of interactions via the aforementioned Luttinger parameter $K$, Eq.~\eqref{eq:K}.

Here, we present results by simulating time evolution in the dimensionless volume $l = 10$. Finite size effects from excitations travelling around the volume limit the evolution time to $m_1t < l$. However, lower volumes are less computationally demanding, and we also find that the time range allowed by this choice is suitable for a detailed comparison of the two methods. We also performed a few computations in larger volumes up to $l = 18$ and found that all the conclusions drawn in this paper remained unchanged.

Below, we consider two different types of quantum quenches. To demonstrate the power of the MSTHA, we first focus on weak quenches inducing a small energy density in Sec.~\ref{QP-quenches}. Here the initial state is close to the quantum pendulum ground state associated with the post-quench Hamiltonian, such that the basis used in MSTHA is well-suited for representing the time evolution of the state, in contrast to previous implementations of TCSA. As a result, MSTHA yields reliable, well-converged results for a wide range of interaction strengths, from hard core repulsion to the experimentally relevant weakly interacting limit, substantially extending the quench protocols accessible within the framework of Hamiltonian truncation. 

For completeness, in Sec.~\ref{FB-quenches}, we revisit strong quenches from the ground state of the unperturbed $(\lambda=0)$ free bosons, i.e., two decoupled one-dimensional quasi condensates in the experimental setup, to finite $\lambda$ / Josephson coupling. These protocols were already studied relying on previous TCSA implementations~\cite{Horvath2019}, formulated in terms of the eigenstates of the massless free boson limit, a natural choice for representing the initial state. For these strong quenches, both TCSA and MSTHA suffer from similar limitations, yielding well-converged results for strong interactions but breaking down in the experimentally relevant weakly interacting regime. Nevertheless, we find that MSTHA still shows improved convergence properties.

\subsection{Quantum quenches starting from the quantum pendulum ground state}\label{QP-quenches}

\begin{figure*}[t]
    \centering
    \includegraphics[width=\linewidth]{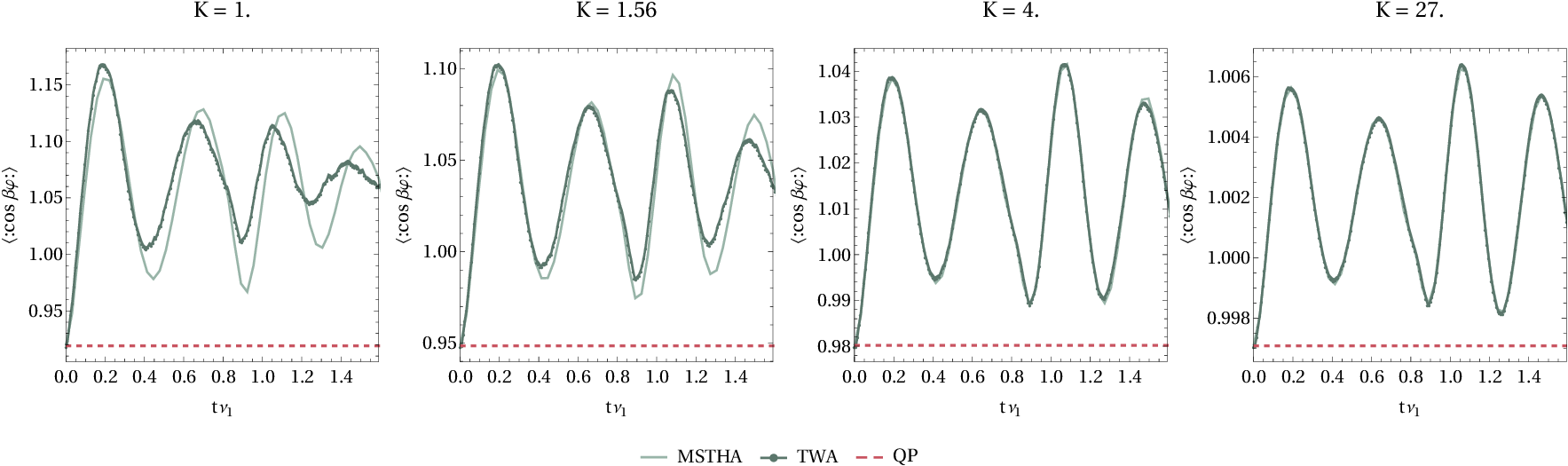}\\
    \includegraphics[width=\linewidth]{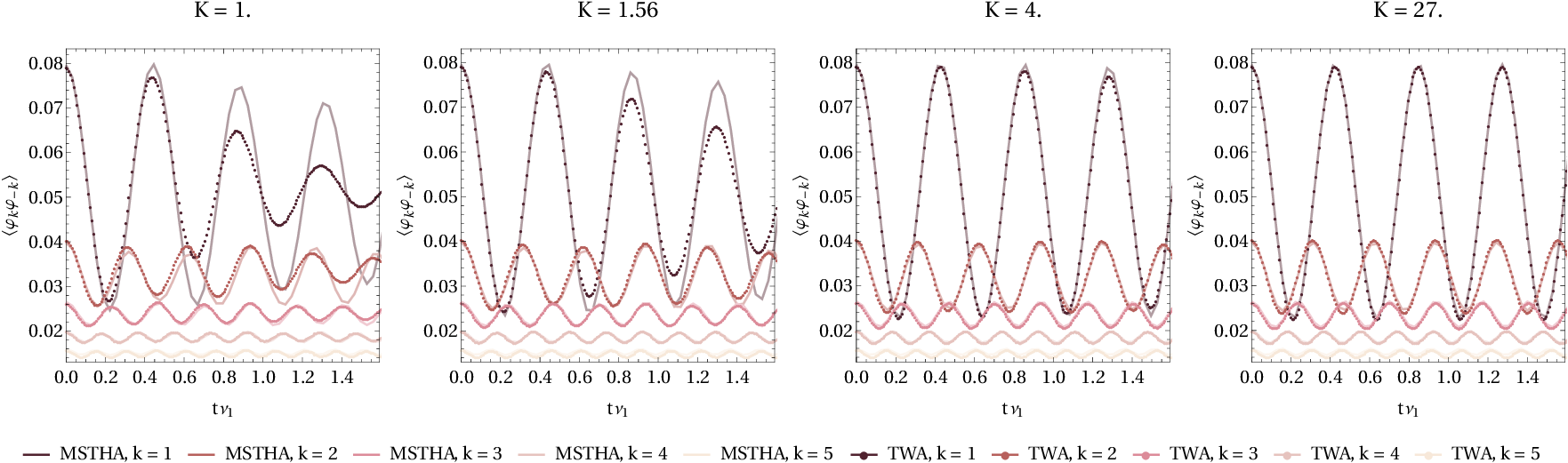}
    \caption{The time-dependent expectation value of $:\cos\beta\hat{\varphi}:$ (top row) and the phase-phase correlator $\expval{\hat{\varphi}_k\hat{\varphi}_{-k}}$ (bottom row) for various values of $K$ for dimensionless volume $l = 10$, starting from the initial state (\ref{qp-init-state}). Joined markers correspond to TWA, while solid lines show the MSTHA results. For the two larger values of $K$, the difference between the two approximations is entirely invisible. The dashed red line corresponds to the (numerically) exact solution of the zero-mode quantum pendulum dynamics. }
    \label{fig:cos-qpgs}
\end{figure*}

Here, we consider weak quantum quenches starting from the quantum pendulum ground state, corresponding to a small injected energy density. More precisely, the initial state corresponds to the zero mode being in its ground state, 
\begin{equation}
    \ket{\Psi_\text{QP}} = \ket{n = 0},\label{qp-init-state}
\end{equation}
while all other modes are in the ground state of the respective oscillator.
The time evolution can be interpreted by suddenly switching on the coupling between the zero-mode pendulum and the non-zero modes corresponding to phononic excitations. This scenario is expected to be optimal for the MSTHA since the implementation uses the pendulum eigenstate basis for the zero-modes, and the energy injected by the quench into the system is small, increasing the reliability of the truncated approximation.

We also compare the MSTHA to the TWA approach. To this end, the Wigner function of the initial state can be decomposed into a product of the part corresponding to the zero mode and the one coming from the oscillator modes. The zero mode part can be obtained simply from the numerically computed ground state wave function of the pendulum Hamiltonian \eqref{qp-ham}, 
\begin{align}
    W_0\left(\varphi_0, \pi_0\right)=\frac{1}{2\pi}\int_{-\pi}^\pi &d\varphi_0'\braket{\varphi_0-\varphi_0'/2}{n=0} 
    \nonumber\\
    &\braket{n=0}{\varphi_0+\varphi_0'/2}e^{-iN\varphi_0'\pi_0)}\,.
\end{align}
Here, the matrix element $\braket{\varphi}{n=0}$ is just the ground state wave function of the pendulum \eqref{qp-ham} in position space. For the non-zero modes, the Wigner function takes the form of a simple Gaussian\cite{Horvath2019}
\begin{align}
    W_\text{osc}=&\prod\limits_{k>0}\frac{4}{\pi^2}
    \exp\left\{
    -\sigma_k^2\varphi_k\varphi_{-k}
    -\frac{4\pi_k\pi_{-k}}{\sigma_k^2}
    \right\}\nonumber\\
    &\sigma_k^2=4 N\sin\frac{\pi k}{N}\to 4\pi k\quad\text{for } N\to\infty\,.
    \label{eq:Wosc}
\end{align}
We note that the quench protocol discussed here, coupling a quantum pendulum to a bath of massless modes, does not have direct experimental relevance. From the experimental point of view, a potentially more realistic state would have the non-zero modes in the ground state of appropriate massive oscillator modes. The present choice, where they are described as gapless modes of the conformal boson, eventually overestimates their contributions and is motivated by two considerations. First, it leads to a technical simplification since the above state has a simpler representation in terms of the MSTHA. Second, one of our goals is to gauge whether these modes play a significant role in the dynamics and determine how much they alter the quantum pendulum dynamics. As mentioned above, this step is crucial for finding the simplest theoretical description of the experiments, where the time evolution is potentially affected by many additional degrees of freedom, including the symmetric and transverse modes. To establish the relevant degrees of freedom, it is, therefore, acceptable to consider a slightly modified quench protocol that overestimates the effects of finite momentum modes., rendering the gapless nature of these modes secondary. The implementation of massive modes lies outside the scope of the paper.

Fig. \ref{fig:cos-qpgs} displays the time evolution of the expectation value of the cosine of the phase field and the phase-phase correlator starting from the initial state (\ref{qp-init-state}) for several interaction strengths $K$ and for a dimensionless volume $l = 10$ as computed by the MSTHA and the TWA. The MSTHA data is computed using truncation values of $n_{\text{max}} = 9, 7, 11$ and $7$ and $\ell_{\text{max}} = 20, 20, 20$ and $24$, corresponding to Hilbert space dimensions of 32247, 25081, 39413 and 88536 for $K = 1, 1.56, 4$ and $27$, respectively. The largest value is eventually the one directly relevant in the experimental context.

The MSTHA results converge well with the truncation and can be considered numerically exact. Since the quantum pendulum is initially in its ground state, a small number of zero mode basis states is enough for the MSTHA to converge. In contrast, the TWA does not allow for a reliable estimate of its accuracy; however, due to the numerically exact nature of the MSTHA, the deviation of the TWA from the MSTHA results can be considered the error involved in the TWA approximation. 

We can see that the two methods agree well for large values of $K$, where quantum effects are expected to be small, which is reasonable given the TWA's semiclassical nature. However, for smaller values of $K$ where quantum fluctuations are enhanced, the TWA differs from the numerically exact results of the MSTHA. This is also expected in light of the initial state (\ref{qp-init-state}): insufficient energy is injected into the system to accommodate higher occupation numbers in the oscillator modes, enhancing the inherently quantum nature of their dynamics. The difference in the time evolution is that the TWA overestimates the dephasing of the condensates, as shown by the results in Fig. \ref{fig:cos-qpgs}.

For the case studied in this section, truncating the Hamiltonian to the zero-mode part \eqref{qp-ham} results in trivial time evolution since the initial state is its eigenstate. Therefore, the strength of interaction between the zero-mode pendulum and the phononic modes can be deduced from the dynamic range of the cosine expectation value in  Fig. \ref{fig:cos-qpgs}, which decreases substantially for large $K$ and becomes very small at the experimentally relevant value $K=27$, indicating that the zero-mode is very weakly coupled to the phononic excitations.

\subsection{Quantum quenches starting from the free massless boson vacuum}\label{FB-quenches}

\begin{figure*}[t]
    \centering
    \includegraphics[width=\linewidth]{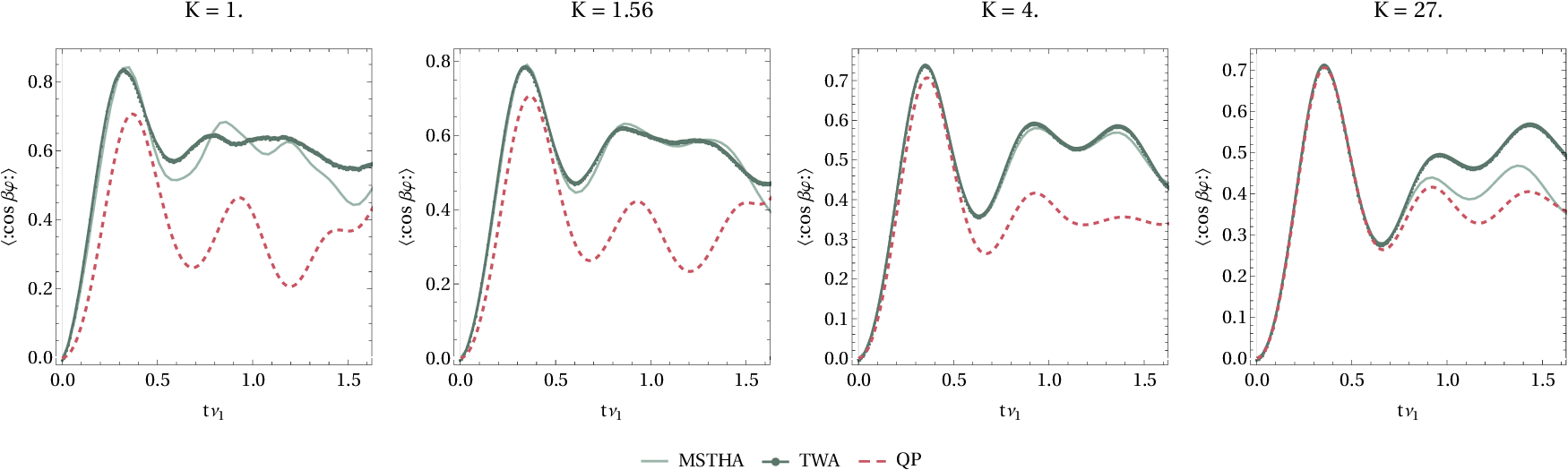}\\
    \includegraphics[width=\linewidth]{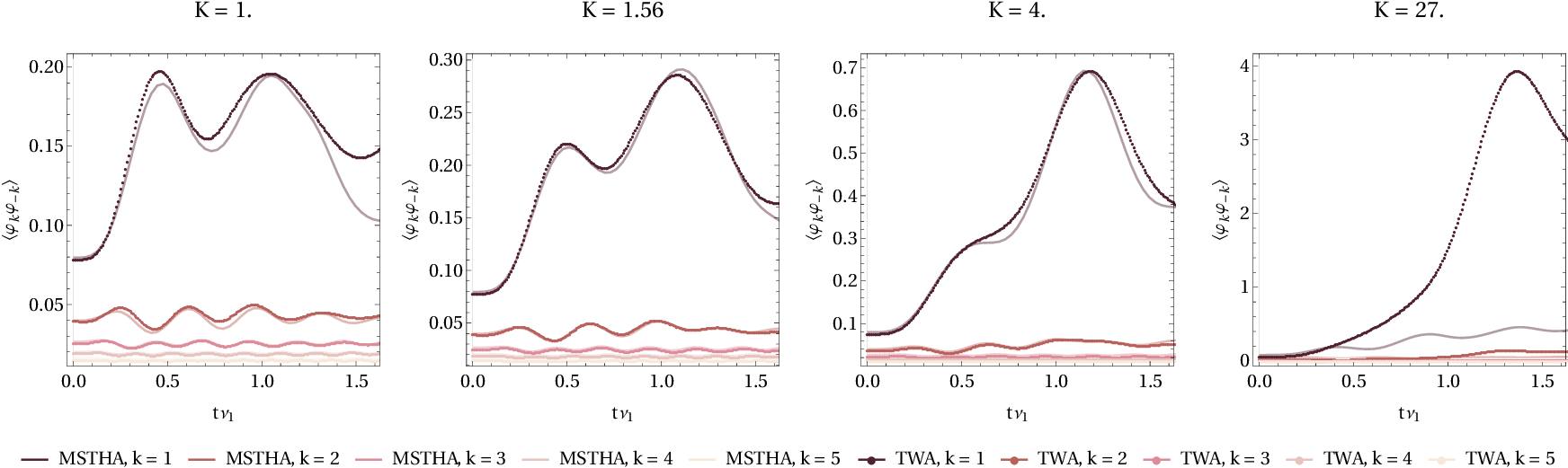}
    \caption{The time-dependent expectation value of $:\cos\beta\hat{\varphi}:$ (top row) and the phase-phase correlator $\expval{\hat{\varphi}_k\hat{\varphi}_{-k}}$ (bottom row) for various values of $K$ for dimensionless volume $l = 10$, starting from the state (\ref{fb-init-state}). Joined markers correspond to TWA, while solid green lines show the MSTHA results. The dashed red lines correspond to the (numerically) exact solution of the zero-mode quantum pendulum dynamics.}
    \label{fig:cos-fbgs}
\end{figure*}

Here, we consider the quenches from the ground state of the unperturbed ($\lambda=0$) free boson,
\begin{equation}
    \ket{\Psi_\text{FB}} = \ket{\nu = 0}\label{fb-init-state}
\end{equation}
which is more directly relevant to the experiment than the previous one. It can be realised by cooling the atoms in the presence of a large barrier to obtain two uncoupled identical condensates and then introducing Josephson tunnel-coupling via lowering the barrier to achieve a desired finite $\lambda$. The time evolution starting from this initial state was previously studied using TWA and TCSA\cite{Horvath2019}. The drawback of that study was that the original version of TCSA (sketched in Subsection \ref{subsec:TCSA}) was limited to rather small values $K$ far away from the experimentally realised weak coupling regime. 

The state (\ref{fb-init-state}) can be easily implemented in the MSTHA by expanding the plane wave state $\ket{\nu=0}$ in the eigenstates of the zero-mode pendulum Hamiltonian (\ref{qp-ham}):
\begin{equation}
    \ket{\Psi_\text{FB}} = \ket{\nu = 0} = \sum_{n = -N}^{N}C_n\ket{n}
\end{equation}
Accurately representing this state requires more vectors for larger values of $K$, which foreshadows that MSTHA has difficulties capturing the time evolution for large values of $K$. While this is similar to the original TCSA\cite{Horvath2019}, we still find that the mini-superspace representation substantially improves the situation.

In the TWA, the Wigner distribution again factorises into a zero-mode part 
\begin{equation}
    W_0\{\varphi_0, \pi_0\} = \frac{\theta(\varphi_0+\pi)\theta(\pi-\varphi_0)}{2\pi}\delta_{\pi_0,0}
\end{equation}
with the oscillator part identical to \eqref{eq:Wosc}. The above zero-mode part corresponds to a uniform distribution of initial phases $\varphi_0$ in the range $\left[-\pi, \pi\right]$ together with a definite value  $\pi_0 = 0$.

The results of the TWA and MSTHA simulations for the time evolution of $\expval{:\cos\beta\hat{\varphi}:}$ and $\expval{\hat{\varphi}_k\hat{\varphi}_{-k}}$ following a quantum quench from the initial state (\ref{fb-init-state}) are shown in Fig. \ref{fig:cos-fbgs}. Simulations were performed for $l = 10$ and several Luttinger parameters $K$. As noted above, contrary to the case where the system is initialised in the ground state of the quantum pendulum, the conformal vacuum (\ref{fb-init-state}) spans a large subspace of the quantum pendulum eigenbasis, requiring larger cutoff values in the mini-superspace:  $n_{\text{max}} = 11, 17, 35$ and $225$ for $K = 1, 1.56, 4$ and $27$, respectively (as before, the largest value is the one relevant for the experimental realisation). The zero-mode cutoff values are chosen so the dynamics remains unchanged by increasing the cutoff $n_{\text{max}}$.  For the oscillator modes, we used the truncations $\ell_{\text{max}} = 26, 20, 28$ and $20$, corresponding to 251339, 60911, 2521750 and 806175 for $K = 1, 1.56, 4$ and $27$, respectively. For the couplings $K = 1$ and $1.56$, the simulations converged with high accuracy, and in the latter case, they also matched the TWA results. However, for $K = 4$ the MSTHA simulations involving larger Luttinger parameters converged less well. Nevertheless, we found that the results matched the TWA results very well. For $K=27$, MSTHA failed to converge for the accessible truncation levels, pointing to the need to include higher excitations in the oscillation modes, making the use of MSTHA computationally extremely demanding.

Again, the TWA fails to describe the time evolution for the strongly interacting regime, as evidenced by its deviation from the MSTHA, which can be considered numerically exact. Nevertheless, TWA shows improved performance due to the high energy density induced by the quench. In particular, the TWA becomes much better for larger $K$, and in fact very accurate for $K\gtrsim 2$, making it a reliable description in the experimental regime. We also note that the TWA gives very good results even for $K = 1.56$, indicated by the minimal disagreement with the MSTHA data. This contrasts with the quenches from the quantum pendulum ground state, where the small energy injected in the quench forbids the accumulation of large occupations in the oscillator modes, amplifying the difference between the quantum and the semiclassical dynamics. In quenches starting from the ground state of the massless free boson, the system is initialised in a very highly excited state, as indicated by the large values of $n_\text{max}$ required to represent the time-evolving state. When the interaction between the zero-mode pendulum and the phononic modes is switched on at time $t = 0$, a large amount of energy is transferred into the oscillator modes, resulting in mode occupation numbers seen in Fig. \ref{fig:cos-fbgs}, which are much higher compared to those in Fig. \ref{fig:cos-qpgs}. The occupation of these modes grows with $K$, and their presence decoheres the zero mode dynamics, which, together with the suppression of quantum fluctuations, accounts for the good agreement with the semiclassical TWA results. However, as $K$ decreases, the effects of quantum fluctuations grow, and the occupation numbers of the oscillator modes decrease, which explains the growing deviation between the semiclassical TWA and the full quantum dynamics obtained from the MSTHA.

Similarly to the previous case, the red dashed lines in Fig. \ref{fig:cos-fbgs} show time evolution considering only the zero-mode dynamics governed by the quantum pendulum Hamiltonian \eqref{qp-ham}. Again, we find that the zero mode dominates the dynamics for large $K$; however, even at the very large $K$, which is characteristic of the experiment, the oscillating modes are seen to influence the dynamics substantially as time progresses. This is fully consistent with the energy transfer to the oscillating modes, which leads to a substantial increase in their occupation number, counteracting the effect of their weaker coupling to the zero mode.  

\section{Conclusions}\label{sec:conclusions}

This work investigated the non-equilibrium time evolution induced by quantum quenches in the sine-Gordon model. Besides being a paradigmatic example of integrable quantum field theories, the sine-Gordon model also describes the low-energy dynamics of two Josephson-coupled one-dimensional bosonic quasi-condensates.\cite{Hofferberth_2007,Langen_2013,Gring_2012,Tajik2022, Bouchoule2005} However, the experimental system has many additional degrees of freedom, which are not accounted for in the sine-Gordon description. Simulating the physical system realised in the experiment is still an open question, and progress requires the identification of the relevant degrees of freedom.

Motivated by the fact that the coupling between the zero and non-zero modes of the sine-Gordon field is weak in the experimentally available parameter range, a naturally occurring question is the importance of non-zero modes for the dynamics. To address this issue, we introduced the mini-superspace-based truncated Hamiltonian approximation (MSTHA), an improvement of the truncated conformal space approach (TCSA) used in earlier studies\cite{Feverati_1998,Horvath2019}. It consists of solving the zero-mode dynamics in a numerically exact way and then including the non-linearly interacting phononic modes. Apart from making the distinction between the zero and non-zero modes explicit, it also efficiently improves the previous versions of the THA, allowing for the simulations in the weakly interacting regime closer to the experiments. In addition, we used the semiclassical truncated Wigner approximation\cite{Polkovnikov2003, POLKOVNIKOV2010} (TWA) as an alternative approach, a simple and wide-spread method that has been applied for various sine-Gordon quenches. Comparison to MSTHA allows for studying the accuracy and limitations of the TWA, for which accuracy is hard to control directly.

We considered time evolution from two classes of initial states, corresponding to small and large energy densities, respectively. We find that for the mild quench protocol, starting in the ground state of the quantum pendulum, the MSTHA yields essentially (numerically) exact results regardless of the Luttinger parameter $K$, even in the weakly interacting limit relevant to the experiments, a region inaccessible by previous implementations of the THA\cite{Horvath2019}. For the stronger quenches initiated in the ground state of the free massless boson, the MSTHA results converge for smaller $K$, corresponding to strong inter-mode interactions, but become less reliable with increasing $K$, when the coupling between the modes is weak.

We established that (as generally expected) the TWA performs well in the weakly interacting regime, even for the mild quench protocol, indicated by the virtually non-existent difference from the MSTHA results. However, this difference grows as the strength of the interaction increases, leading to the breakdown of the TWA close to $K = 1$, corresponding to hard-core repulsion between atoms. While this trend remains unchanged for stronger quenches as well, it is found that the reliability of the TWA increases with the strength of the quench, pushing its breakdown to smaller values of $K$ compared to mild quenches. This latter effect is intuitively expected since the TWA is a semiclassical approximation, which is expected to improve with higher excitations in the modes.

Our findings establish the TWA and MSTHA as powerful numerical methods for studying non-equilibrium dynamics in the sine-Gordon model, depending on the initial state and strength of the inter-mode interaction $K$. For weak quenches or strong quenches for strong interactions (large $K$), the MSTHA can provide reliable results for the dynamics, while for strong quenches or weak interactions, the TWA proves reliable for studying the time evolution. Overall, our results establish the TWA and MSTHA as powerful complementary approaches for studying non-equilibrium time evolution in the sine-Gordon model in the weakly interacting parameter range accessible in the experiments, with the choice of method dependent on the initial energy density of the system.

Moreover, we find that the effect of the nonzero modes, a.k.a. the phononic degrees of freedom, diminishes when the interaction becomes weaker (i.e., for large $K$) and has a limited effect on the time evolution for mild quenches. For stronger quenches, the contribution of the phononic modes becomes weaker for the initial transient; however, even in the experimentally relevant large $K$ regime, it eventually appears when the occupation number of the phononic modes becomes large.

\begin{acknowledgments}
We thank S. Erne and J. Schmiedmayer for useful discussions and D. Horváth for sharing his TCSA results to verify our numerics. This work was supported by the  National Research, Development and Innovation Office (NKFIH) through the OTKA Grant ANN 142584. DSz was also partially supported by the National Research Development and Innovation Office of Hungary via the scholarship ÚNKP-22-3-II-BME-30, while GT was partially supported by the Quantum Information National Laboratory of Hungary (Grant No. 2022-2.1.1-NL-2022-00004). I.L. acknowledges support from the Gordon and Betty Moore Foundation through Grant GBMF8690 to UCSB and the National Science Foundation under Grant No. NSF PHY-1748958.
\end{acknowledgments}

\appendix
\section{Matrix elements on the CFT basis}\label{appendix-matelements}
For the practical evaluation of matrix elements of the exponential operators in the computational basis, time is continued to Euclidean signature by setting $t=i\tau$, and then the resulting space-time cylinder is mapped on the conformal plane of variable $z$ using\cite{Horvath2022}
\begin{equation}
    z=\exp\left\{\frac{2\pi}{L}(\tau-i x)\right\}\quad,\quad
    \bar{z}=\exp\left\{\frac{2\pi}{L}(\tau+i x)\right\}\,.
\end{equation}
The exponential operator on the cylinder is related to the one defined on the plane by 
\begin{equation}
:e^{i\beta\nu\hat{\varphi}}:^\textrm{cyl} = \left(\frac{2\pi|z|}{L}\right)^{2\Delta_\nu}:e^{i\beta\nu\hat{\varphi}}:^\textrm{pl}
\end{equation}
with
\begin{equation}
    \Delta_\nu = \frac{\nu^2\beta^2}{8\pi}
\end{equation}
Therefore, the matrix elements of the integrated exponential operator can then be computed as
\begin{equation}
\begin{split}
&\int\limits_0^L dx \matrixelement{\Psi'}{:\exp\left\{i\nu\beta\hat{\varphi}(0,x)\right\}:^\textrm{cyl}}{\Psi}
\\
&=L\left(\frac{2\pi}{L}\right)^{2-2\Delta_\nu}\matrixelement{\Psi'}{:\exp\left\{i\nu\beta\hat{\varphi}(1,1)\right\}:^\textrm{pl}}{\Psi}\delta_{s_\Psi s_{\Psi'}}\, .
\end{split}
\end{equation}
Implementation of the above matrix element requires the computation of
\begin{equation}
    \matrixelement{\Psi'}{:\exp\left\{i\mu\beta\hat{\varphi}(1,1)\right\}:^\textrm{pl}}{\Psi}\,, \label{vertex-matelem}
\end{equation}
which is a straightforward task described in detail in previous works\cite{Horvath2019, Horvath2022}. 

\subsection{Pendulum quantum mechanics}

Implementation of the mini-superspace for the sine-Gordon model requires the construction of the quantum pendulum Hamiltonian
\begin{equation}
    \hat{H}_\text{QP} = \frac{1}{2L}\hat{\pi}_0^2 -\lambda L \left(\frac{2\pi}{L}\right)^{2\Delta}\cos(\beta\hat{\varphi}_0)\; \label{qp-ham-app}.
\end{equation}
The free part
\begin{equation}
    \hat{H}_\text{FQM} = \frac{1}{2L}\hat{\pi}_0^2
\end{equation}
admits solutions $\ket{\nu}$ in the form of plane waves:
\begin{align}
    &\ket{\nu} = \sqrt{\frac{\beta}{2\pi}}e^{i\beta\nu\varphi_0}\\
    &\frac{1}{2L}\hat{\pi}_0^2\ket{\nu} = \frac{(\nu\beta)^2}{2L}\ket{\nu}\; ,
\end{align}
with the canonical momentum operator given in coordinate representation as
\begin{equation}
    \hat{\pi}_0 = \frac{1}{i}\partial_{\varphi_{0}}\; .
\end{equation}
The states $\{\ket{\nu}\}$ are created by the exponential operators from the vacuum
\begin{equation}
    \ket{\nu} = e^{i\beta\nu\hat{\varphi}_0}\ket{0}\; ,\quad\ket{0} = \frac{\beta}{2\pi}
\end{equation} 
and therefore, the zero-mode exponential operators act as ladder operators on the plane wave basis:
\begin{equation}
    e^{\pm i\mu\beta\hat{\varphi}_0}\ket{\nu} = \ket{\nu\pm\mu}\; .
\end{equation}
Employing a simple truncation of the plane wave basis by only keeping states $\{\ket{\nu}\}, \nu\in [-\nu_\text{max}, \nu_\text{max}]$ results in a representation of the operators $\hat{\pi}_0^2$ and $\exp{\pm i\beta\nu\hat{\varphi}_0}$ by finite matrices, with the Hamiltonian (\ref{qp-ham-app}) becoming a tridiagonal matrix.

Numerical diagonalisation of the finite matrix of (\ref{qp-ham-app}) is straightforward and leads to the spectrum of the quantum pendulum
\begin{equation}
    \hat{H}_\text{QP}\ket{n} = \varepsilon_n\ket{n}\; ,
\end{equation}
which we cross-checked by numerically solving the coordinate space Schrödinger equation with the shooting method.

\section{Low-energy description of a pair of coupled bosonic quasi-condensates}\label{app:condensates}
The Hamiltonian of a one-dimensional bosonic quasi-condensate is given by
\begin{align}\label{ham-QC}
    \hat{H}_\text{QC} = \int dz~\hat{\psi}^{\dagger}(z)\left[-\frac{\hbar}{2m}\partial_z^2 + V(z) - \mu\right]\hat{\psi}(z)\nonumber\\ + \frac{g}{2} \int dz~\hat{\psi}^{\dagger}(z)\hat{\psi}^{\dagger}(z)\hat{\psi}(z)\hat{\psi}(z)
\end{align}
where the $\hat{\psi}$ are bosonic field operators satisfying $[\hat{\psi}(z), \hat{\psi}^{\dagger}(z')] = i\delta (z-z')$, $V(z)$ is a longitudinal trap potential, $\mu$ is the chemical potential and $g$ is some effective one-dimensional interaction coupling. The strength of the interaction is characterised by the parameter
\begin{equation}
    \gamma =  \frac{m g}{\hbar^2 \rho_0}\,,
\end{equation}
where $\rho_0$ is the longitudinal density of atoms. Introducing the bosonisation in terms of density $\hat{\rho}(z)$ and phase $\hat{\theta}(z)$ fields,
\begin{equation}\label{madelung}
    \hat{\psi}(z) = \sqrt{\hat{\rho}(z)}e^{i\hat{\theta}(z)};\quad\hat{\rho} = \rho_0 + \delta\hat{\rho}\,,
\end{equation}
the density fluctuations $\delta\hat{\rho}$ and the phase field $\hat{\theta}$ obey the commutation relations $[\hat{\theta}(z), \delta\hat{\rho}(z')] = i\delta(z-z')$. Substituting (\ref{madelung}) to (\ref{ham-QC}) and expanding to second order in density and phase fluctuations yields a low-energy effective field theory in the form of the Tomonaga-Luttinger-liquid Hamiltonian
\begin{equation}
    \hat{H}_\text{TLL} = \frac{\hbar}{2\pi}\int dz~\left[\nu_N \pi^2 \delta\hat{\rho}^2 + \nu_J (\partial_z\hat{\theta})^2\right].
\end{equation}
Here the density/phase-stiffness $\nu_{N/J}$ can be expressed in terms of the parameters of the condensate,
\begin{equation}
    \nu_J = \frac{\pi\hbar \rho_0}{m},\quad\quad\nu_N = \frac{1}{\pi \hbar}\partial_{\rho_0}\mu \stackrel{\gamma \ll 1}{\approx}\frac{g}{\pi \hbar} 
\end{equation}
Due to the spatial dependence of the background density $\rho_0$ (inherited from the trapping potential $V(z)$), these parameters generally carry a $z$-dependence, which we ignore from now on, focusing on a homogenous system. Introducing the Luttinger parameter $K$ and sound velocity $c$ as
\begin{equation}
    \tilde{K} = \sqrt{\frac{\nu_J}{\nu_N}}, \quad\quad c = \sqrt{\nu_J \nu_N},
\end{equation}
results in the following form the Tomonaga-Luttinger Hamiltonian,
\begin{equation}\label{ham-TLL}
    \hat{H}_\text{TLL} = \frac{\hbar c}{2}\int dz~\left[\frac{\pi}{\tilde{K}} \delta\hat{\rho}^2 + \frac{\tilde{K}}{\pi} (\partial_z\hat{\theta})^2\right].
\end{equation}

\begin{figure*}[]
    \centering
    \includegraphics[width = 0.4\linewidth]{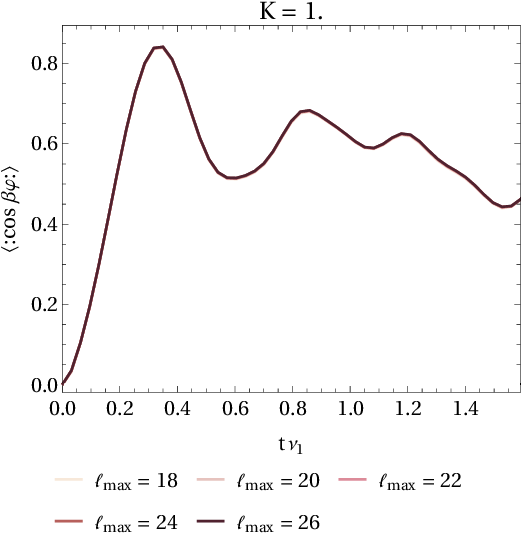}\quad
    \includegraphics[width = 0.4\linewidth]{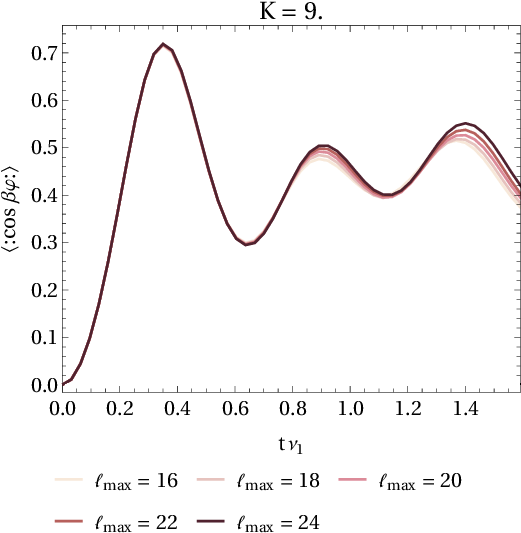}
    \caption{Examples on the convergence of the MSTHA for $l = 10$ and two different values of $K$. The left panel shows the $\expval{:\cos\beta\hat{\varphi}:}$ for $K = 1$ and different level-cutoffs $\ell_{\text{max}}$ where the MSTHA simulations converged. On the right panel simulations corresponding to $K = 9.$ and various $\ell_{\text{max}}$ are shown where the MSTHA failed to converge, resulting in remaining truncation errors in the results.}
    \label{fig:convergence}
\end{figure*}

For a pair of bosonic one-dimensional quasi-condensates loaded into a double-well potential, a finite potential barrier induces a coupling between the condensates through tunnelling, described by the Hamiltonian
\begin{align}
    \hat{H}_\text{QCP} = \sum_{j = 1,2} \int dz~\hat{\psi}_j^{\dagger}(z)\left[-\frac{\hbar}{2m}\partial_z^2 + V(z) - \mu_j\right]\hat{\psi}_j(z)\nonumber\\ + \frac{g}{2} \int dz~\hat{\psi}_j^{\dagger}(z)\hat{\psi}_j^{\dagger}(z)\hat{\psi}_j(z)\hat{\psi}_j(z) - \hbar J \int dz~\left[\hat{\psi}_1^\dagger\hat{\psi}_2 + \hat{\psi}_2^\dagger \hat{\psi}_1\right],
\end{align}
with tunelling amplitude $J$. Setting $\mu_1= \mu_2 = \mu$, introducing bosonisation via
\begin{equation}
    \hat{\psi}_j(z) = \sqrt{\hat{\rho}_j(z)}e^{i\hat{\theta}_j(z)};\quad\hat{\rho} = \rho_0 + \delta\hat{\rho}_j
\end{equation}
and expanding to second order in the fluctuations we arrive at 
\begin{equation}
    \hat{H}_\text{QCP} = \hat{H}_\text{TLL, 1}(\tilde{K})+ \hat{H}_\text{TLL, 2}(\tilde{K}) + \hat{H}_J(\hat{\theta}_1-\hat{\theta}_2)\,,
\end{equation}
where we explicitly indicated the Luttinger parameter. Since the coupling Hamiltonian $\hat{H}_J$ only depends on the relative phase of the two quasi-condensates, it is advantageous to perform a change of variables to common and relative degrees of freedom as
\begin{align*}
     &\delta\hat{\rho}_c = \delta\hat{\rho}_1 + \delta\hat{\rho}_2\\
     &\delta\hat{\rho}_r = \frac{\delta\hat{\rho}_1 - \delta\hat{\rho}_2}{2}\\
     &\hat{\theta}_c = \frac{\hat{\theta}_1 + \hat{\theta}_2}{2}\\
     &\hat{\theta}_r = \hat{\theta}_1 - \hat{\theta}_2.
\end{align*}
The TLL Hamiltonians can be rearranged as
\begin{equation}
    \hat{H}_\text{TLL, 1}(\tilde{K})+ \hat{H}_\text{TLL, 2}(\tilde{K}) = \hat{H}_\text{TLL, c}(K_c)+ \hat{H}_\text{TLL, r}(K_r)
\end{equation}
with $K_c = 2 \tilde{K}$ and $K_r = \tilde{K}/2$, while expanding to second order in density fluctuations we obtain
\begin{align}
    \hat{H}_J &= -\hbar J\int dz~\left[2\rho_0 + \delta\hat{\rho}_c\right](\cos\hat{\theta}_r - 1) + \frac{\hbar J}{\rho_0}\delta\hat{\rho}_r^2\cos\hat{\theta}_r \nonumber\\
    &\approx -2\hbar J\rho_0\int dz~\cos\hat{\theta}_r\,.
\end{align}
Here, the second line was obtained by neglecting the density fluctuations, resulting in a decoupling of common and relative degrees of freedom: 
\begin{align}
    &\hat{H}_c = \frac{\hbar c}{2}\int dz~\left[\frac{\pi}{2\tilde{K}} \delta\hat{\rho}_c^2 + \frac{2\tilde{K}}{\pi} (\partial_z\hat{\theta}_c)^2\right]\label{eq:common-cond}\\
    &\hat{H}_r = \frac{\hbar c}{2}\int dz~\left[\frac{2\pi}{\tilde{K}} \delta\hat{\rho}_r^2 + \frac{\tilde{K}}{2\pi} (\partial_z\hat{\theta}_r)^2\right] - 2\hbar J \rho_0\int dz~\cos\hat{\theta}_r\,\label{eq:rel-cond}
\end{align}
showing that the relative phase field obeys sine-Gordon dynamics. To further simplify the Hamiltonian, we choose units in which $\hbar = c = 1$ and introduce a  boson field and its canonical momentum defined as
\begin{equation}
\hat{\varphi}=\beta^{-1}\hat{\theta}_r\quad,\quad 
\hat{\pi}=\beta\delta \hat{\rho}_r
\end{equation}
where  
\begin{equation}
    \beta = \sqrt{\frac{2\pi}{\tilde{K}}}\,,
\end{equation}
which leads to the usual form of the sine-Gordon Hamiltonian
\begin{equation}
    \hat{H}_{\text{sG}} = \frac{1}{2}\int dz\left[\hat{\pi}^2 + (\partial_z\hat{\varphi})^2\right] - \lambda\int dz: \cos\beta\hat{\varphi}:\,.
\end{equation}
We note that the normal ordering results in a redefinition of the coupling $\lambda$ and accounts for its anomalous dimension $\Delta=\beta^2/8\pi$, manifesting in $\lambda$ having units of $[\text{energy}]^{2-2\Delta}$.

We also note that in the main text, we consider only the physics of the relative degrees of freedom, and so we drop the subscript of $K_r$ and refer to the relative Luttinger parameter simply as $K$.

\section{Examples on the convergence of MSTHA}\label{app:mstha_convergence}

This section contains representative data illustrating the convergence of the MSTHA. In Fig. \ref{fig:convergence}, the time evolution of the cosine of the phase field is shown as computed by the MSTHA starting from the massless free boson ground state (\ref{fb-init-state}). It is apparent that for small values of $K$, the MSTHA quickly converges, whereas for larger $K$ the convergence is much slower, as displayed in Fig. \ref{fig:convergence}.

\bibliographystyle{utphys.bst}
\bibliography{mstha.bib}

\end{document}